\documentclass{IEEEtran}
\usepackage[utf8]{inputenc}
\usepackage[T1]{fontenc}
\usepackage{comment}
\usepackage{amsmath}
\usepackage{amsfonts}
\usepackage[english]{babel}
\usepackage{amsthm}
\usepackage{amssymb}
\usepackage{graphicx}
\usepackage{latexsym}
\usepackage{algorithm}
\usepackage{algpseudocode}
\usepackage{lipsum}
\usepackage{mathrsfs}
\usepackage{enumitem}
\usepackage{tikz, pgfplots}
\usepackage{eso-pic,xcolor}
\usepackage{overpic, scalerel}
\pgfplotsset{compat=1.18}
\usetikzlibrary{shapes,plotmarks}
\IEEEoverridecommandlockouts
\usepackage{subcaption}
\usepackage{multicol}
\usepackage{hyperref}
\usepackage[hyphenbreaks]{breakurl}
\usepackage{cleveref}
\usepackage{nicematrix}
\usepackage{cite}
\usepackage{booktabs} 
\usetikzlibrary{patterns}        
\newcommand{\BK}{\mathcal{K}}
\newcommand{\customref}[2]{\hyperref[#1]{#2}}

\definecolor{customGreen}{rgb}{0, 0.502, 0}
\definecolor{customyellow}{RGB}{235,178,2}
\definecolor{Egreen}{RGB}{65,136,35}

\newcommand{\smallK}{\scaleto{K=2}{4.5pt}}

\newcommand{\e}{\mathbb{E}}
\newcommand{\se}{\scaleto{rel,k}{4.5pt}}

\newcommand{\tot}{\scaleto{\mathrm{tot}}{4.0pt}}

\newcommand{\RIS}{\scaleto{\mathrm{RIS}}{4pt}}

\newcommand\scalemath[2]{\scalebox{#1}{\mbox{\ensuremath{\displaystyle #2}}}}

\usepackage[figurename=Fig.]{caption}
\newcommand{\GC}{\scaleto{\mathrm{sum}}{3pt}}
\newcommand{\PIN}{\scaleto{\mathrm{Pin}}{4pt}}

\newcommand{\captionwidth}{\captionsetup{width=\linewidth}}
\usepackage{physics}

\title{Pinching Antenna Systems versus Reconfigurable Intelligent Surfaces in mmWave}

\author{{Mostafa Samy,~\IEEEmembership{Graduate Student Member,~IEEE}, Hayder Al-Hraishawi,~\IEEEmembership{Senior Member,~IEEE},\\ Madyan Alsenwi,~\IEEEmembership{Member,~IEEE}, Abuzar B. M. Adam,~\IEEEmembership{Member,~IEEE},\\ Symeon Chatzinotas,~\IEEEmembership{Fellow,~IEEE}, and Björn Otteresten,~\IEEEmembership{Fellow,~IEEE}\vspace{-6mm}}\\
\thanks{M. Samy, M. Alsenwi, A. B. M. Adam, and B. Otteresten are with the Interdisciplinary Centre for Security, Reliability and Trust (SnT), University of Luxembourg, Luxembourg. \\
H. Al-Hraishawi is with the Department of Electrical Engineering, University of South Florida, Tampa, FL 33620 USA.\\
S. Chatzinotas is with SnT, University of Luxembourg, 1855 Luxembourg City, Luxembourg and with College of Electronics \& Information, Kyung Hee University, Yongin-si, 17104, Korea.\\
Corresponding author: \emph{Mostafa Samy (mostafa.samy@uni.lu)}.}
\thanks{This work was supported by the Luxembourg  National Research Fund (FNR) through the AFR Project CEP-MBD-CRIS, Grant reference 17974844.}}

\begin{document}
\pagenumbering{arabic}
\maketitle
\bstctlcite{IEEEexample:BSTcontrol}
\begin{abstract} 
Flexible and intelligent antenna designs, such as pinching antenna systems and reconfigurable intelligent surfaces (RIS), have gained extensive research attention due to their potential to enhance the wireless channels. This letter, for the first time, presents a comparative study between the emerging pinching antenna systems and RIS in millimeter wave (mmWave) bands. Our results reveal that RIS requires an extremely large number of elements (in the order of $10^4$) to outperform pinching antenna systems in terms of spectral efficiency, which severely impact the energy efficiency performance of RIS. Moreover, pinching antenna systems demonstrate greater robustness against hardware impairments and severe path loss typically encountered in high-frequency mmWave bands.
\end{abstract}
\section{Introduction}
The growing demand for ultra-high data rates for next-generation $6$G wireless networks is driving the exploration of spectrum resources, in millimeter-wave (mmWave) bands \cite{mmwave1}. Large bandwidths are available in mmWave and they are well-suited for supporting the increasing number of intelligent devices and latency-sensitive applications that require high data rates \cite{mmwave_RIS}. 
However, one of the fundamental challenges of mmWave
communication is the severe path loss and susceptibility to blockages \cite{mmwave2}.

In light of this, reconfigurable intelligent surfaces (RIS)
are envisioned to overcome link blockage in mmWave
by reflecting signals around obstacles and providing virtual line-of-sight (LoS) connection between transceivers \cite{mmwave_RIS}. The RIS is typically comprised of a large number of low-cost, reconfigurable passive elements, where each element can manipulate the phase shift of an incident electromagnetic wave in a controlled manner, thus enabling a ‘smart’, i.e., programmable, wireless environment \cite{Basar2019}. Despite such potential, RIS inherently suffers from double large-scale attenuation due to its passive nature, which becomes more severe at higher carrier frequencies such as mmWave, where wavelengths are shorter.

Recently, a new flexible-antenna technology, termed pinching antennas,
has been proposed, a more promising solution to combat the large scale path-loss, at higher propagation frequencies \cite{suzuki2022pinching}. Specifically, the location of pinching antennas can be flexibly adjusted along a waveguide, enabling the creation or enhancement of LoS links \cite{pinching_sur}. This can be realized by applying small dielectric particles on a waveguide to operate as leaky-wave antennas. Moreover, dielectric waveguides offer significantly lower propagation loss in high-frequency bands compared to free-space path loss. For instance, a high-purity Teflon-based dielectric waveguide exhibits a propagation loss of approximately $0.08$ dB/m at 28
GHz, whereas the free-space path loss is around $40$ dB/m \cite{pin_per}.

In this direction, important research efforts have been made to assess the benefits of pinching antenna systems across various communication scenarios \cite{downlink_pin_max,pin_per,noma_pin,noma_pin_up,analysis_pin,ISAC_pin}. 
For instance, 
the work in \cite{pin_per} theoretically characterized the ergodic rate achieved by pinching antennas and demonstrated their ability to deliver enhanced spectral performance compared to conventional fixed-position antenna systems. In \cite{downlink_pin_max}, the authors 
studied the downlink rate maximization problem in a pinching-antenna system, where $N$ pinching antennas are deployed on a waveguide to serve a single-antenna user.
The authors in \cite{noma_pin} and \cite{noma_pin_up} optimized the locations of pinching antennas in non-orthogonal multiple access (NOMA) downlink and uplink scenarios, respectively, where multiple antennas are activated along the dielectric waveguide to serve multiple users.
Reference \cite{ISAC_pin} jointly optimized the pinching antenna placement and user power allocation to enhance communication and sensing performance of integrated sensing and communication (ISAC) systems.

Despite this growing interest in exploring pinching antenna systems across various application scenarios in mmWave bands, to the best of the authors’ knowledge, a comparative analysis with RIS technology has not been explored yet.
As the first study to present such comparison, the primary goal of
this letter is to identify the advantages and challenges
associated with both technologies. Furthermore, hardware impairments for pinching antennas and RIS are incorporated to ensure a realistic comparison.

\begin{figure*}[t]
\centering
\vspace{-3mm}
\begin{subfigure}[t]{0.48\linewidth}
 \includegraphics[width=\linewidth]{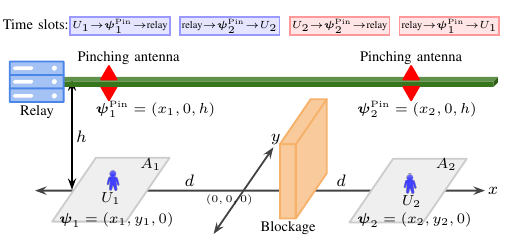}
    \vspace{-6mm}
    \caption{\footnotesize Pinching antenna system: In each time slot, a single pinching antenna is activated to serve $U_k$ during its transmission to or reception from the relay.}
    \label{pin_fig}
\end{subfigure}
    \hfill
\begin{subfigure}[t]{0.48\linewidth}
  \includegraphics[width=\linewidth]{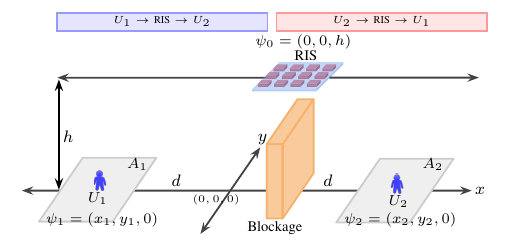}
   \vspace{-6mm}
\caption{\footnotesize RIS-assisted system: $U_1$ transmits in time slot 1, and $U_2$ transmits in time slot 2, where both transmissions are enabled by the RIS.}
   \label{RIS_fig}
\end{subfigure}
    \renewcommand{\figurename}{\footnotesize Fig}
    \vspace{-1mm}
\caption{\footnotesize System model of device-to-device communication for pinching-antenna system and RIS assisted system in mmWave.}
\vspace{-4mm}
\label{sys_model}
\end{figure*}

\section{System Model}\label{sec:sys_mod}
We consider a device-to-device communication scenario, that consists of two single-antenna users, denoted by $U_k$ for $k\in \{1, 2\}$, operating in mmWave band. $U_1$ and $U_2$ aim to exchange data (e.g., file sharing), where the direct link between the two users is not available due to blockage.
To overcome this, we consider two setups to assist the communication, as shown in Fig \ref{sys_model}: A pinching antenna system in Fig \ref{pin_fig}, and RIS-assisted communication system in Fig \ref{RIS_fig}.
We assume the service area of $U_k$ is denoted by $A_k$ and is assumed to be
a square with side length denoted by $L$.
The users are assumed to be uniformly distributed in $A_k$, and $U_k$’s location is denoted by $\boldsymbol{\psi}_k=(x_k,y_k,0)$. The distance from the origin to $A_k$ is denoted by $d$.

\subsection{Pinching Antenna System}
In this setup, a waveguide is placed parallel to the $x$-axis at height $h$ from the origin, and connected to a single RF chain relay, as illustrated in Fig. \ref{pin_fig}. We assume that a pinching antenna can be moved/activated to the location closest
to the served user, and hence, the location of the pinching antenna is $\boldsymbol{\psi}^{\PIN}_k=(x_k,0,h)$ \cite{pin_per}.  In this communication scenario, the transmission is performed in four time slots\footnote{The considered system model is similar to conventional decode-and forward relaying in cooperative communication, however, implemented using a pinching antenna system.}. In the first time slot, $U_1$ transmit the data to the relay using pinching antenna $\boldsymbol{\psi}^{\PIN}_{1}$, and in the second time slot the relay transmit the signal to $U_2$ through pinching antenna $\boldsymbol{\psi}^{\PIN}_{2}$. Similarly, in the third time slot, $U_2$ send its signal to the relay via $\boldsymbol{\psi}^{\PIN}_{2}$, and the relay retransmit the signal to $U_1$ via $\boldsymbol{\psi}^{\PIN}_{1}$.
The reason for not performing the communication directly in two time slots through the channel path $U_1$-$\boldsymbol{\psi}^{\PIN}_{1}$-waveguide-$\boldsymbol{\psi}^{\PIN}_{2}$-$U_2$  will be discussed later.
We assume that the users’ channel state information (CSI) is perfectly
known at the relay, where CSI knowledge is crucial for
positioning the pinching antenna \cite{pin_per}.
Since pinching antennas effectively reduce the transceiver distance, where the likelihood of a line-of-sight (LoS) link increases as this distance shortens, free-space path-loss model is considered \cite{pin_per}. For a fair comparison with the RIS-assisted communication scenario, we assume that the total transmit power allocated per user over its two time slots is $P_k$, meaning both $U_k$ and the relay each transmit with power $\frac{P_k}{2}$.
Based on this, the received SNR of $U_k$ at the relay can be expressed as follows:
\begin{equation}\label{pin_snr_1}
\gamma_{\se}^{\PIN} = \frac{P_k\eta }{2\sigma^2 |\boldsymbol{\psi}^{\PIN}_{k'} - \boldsymbol{\psi}_{k'}|^2}=\frac{P_k\eta }{2\sigma^2 (h^2+y_{k'}^2)},
\end{equation}
where $k' \neq k$ denotes the transmitting user. In \eqref{pin_snr_1},  $\eta = \frac{c^2}{16 \pi^2 f_c^2}$, $c$ denotes the speed of light, $f_c$ is the carrier frequency,  $\sigma^2$ denotes the noise power. Similarly, the received SNR at user $U_k$, when the relay transmits with power $\frac{P_k}{2}$, is given by:
\begin{equation}\label{pin_snr_2}
\gamma_k^{\PIN} = \frac{P_k\eta }{2\sigma^2 |\boldsymbol{\psi}^{\PIN}_k - \boldsymbol{\psi}_{k}|^2}=\frac{P_k\eta }{2\sigma^2 (h^2+y_k^2)}.
\end{equation}
Therefore, the average spectral efficiency for $U_k$ can be given as follows:
\begin{equation}\label{no_clo}
R^{\PIN}_{k} =\mathbb E\qty[\frac{1}{4}\log_2\qty(1+\min\qty(\gamma_{\se}^{\PIN}, \gamma_k^{\PIN}))],
\end{equation}
where the pre-log factor is due to the fact that the communication is performed in four time slots. 
Deriving a closed-form expression for \eqref{no_clo} is challenging, which motivates the use of an upper bound. Assuming symmetric user locations (i.e., $y_1=y_2$), the SNR expressions in \eqref{pin_snr_1} and \eqref{pin_snr_2} become equal. Therefore, $R_k^{\PIN}$ can be obtained from \cite{pin_per} as follows:
\begin{align}\label{ana_pin}
\scalemath{0.9}{R_k^{\PIN} = }&\scalemath{0.9}{\frac{1}{4}\Bigg[\log_2\qty( \frac{L^2}{4} + h^2 + \frac{\eta P_k}{2\sigma^2} )}\nonumber\\
&\scalemath{0.9}{+ \frac{4}{L}  \log_2(e) \sqrt{h^2 + \frac{\eta P_k}{2\sigma^2}} \tan^{-1} \qty( \frac{L}{2 \sqrt{h^2 + \frac{\eta P_k}{2\sigma^2}}} )}\nonumber\\
&\scalemath{0.9}{- \log_2\qty( \frac{L^2}{4} + h^2 )
- \frac{4}{L}  \log_2(e)h  \tan^{-1} \left( \frac{L}{2h} \right)  \Bigg].}
\end{align}
\emph{Remark}:
The system model shown in Fig \ref{pin_fig} suggests that the two users can communicate directly by simultaneously activating the two pinching antennas $\boldsymbol{\psi}^{\PIN}_1$ and $\boldsymbol{\psi}^{\PIN}_2$, i.e., through the channel path $U_1$-$\boldsymbol{\psi}^{\PIN}_{1}$-waveguide-$\boldsymbol{\psi}^{\PIN}_{2}$-$U_2$, where users exchange signals in two orthogonal time slots. At first glance, this communication scheme appears to offer enhanced spectral efficiency, which, however, is not true, as explained in the following.
Considering this scenario, the SNR per user can be given as follows:
\begin{equation}
\gamma_k = \frac{\eta^2P_k }{\sigma^2 |\boldsymbol{\psi}^{\PIN}_1 - \boldsymbol{\psi}_{1}|^2|\boldsymbol{\psi}^{\PIN}_2 - \boldsymbol{\psi}_{2}|^2},
\end{equation}
which shows that the SNR includes a double path loss factor. In other words, the pinching antenna system is acting as a passive device in a manner similar to the RIS, which significantly increases signal attenuation\footnote{Note that when considering hardware impairments, an additional waveguide path loss will be included to the power calculation  of the link budget for this scheme. Hence, the total path loss in this scenario is the product of three components: from $\boldsymbol{\psi}_1$ to $\boldsymbol{\psi}^{\PIN}_1$, waveguide path loss, and from $\boldsymbol{\psi}^{\PIN}_2$ to $\boldsymbol{\psi}_2$.}. We refer to this scheme as passive pinching antenna, which we also evaluate in the numerical section.

\subsection{RIS-Assisted System}
In this alternative setup, we assume a RIS comprising of $M$ reflecting elements, for $M\in \{1,...,M\}$, is deployed at a fixed location in the middle between the two users at hight $h$, i.e., positioned at $\boldsymbol{\psi}_0=(0,0,h)$, as shown in Fig. \ref{RIS_fig}. In this scenario, communication takes place over two time slots, where each user $U_k$ transmits its data to the other through the RIS in one of the slots.
We model the RIS channels in the mmWave band using Rician fading, similar to \cite{sv1,sv2}.
To ensure a fair comparison with the pinching antenna system, a strong line-of-sight scenario (e.g., a Rician factor of 10) will be considered in the numerical results.
In this case, we denote the small-scale fading channels from $U_k$ to the RIS and from the RIS to $U_{k'}$ as \( \mathbf{h} \in \mathbb{C}^{M \times 1} \) and \( \mathbf{g} \in \mathbb{C}^{1 \times M} \), respectively. These are expressed as follows:
\begin{equation}
\mathbf{h}=[h_1,\cdots h_m,\cdots,h_M]^{\mathrm{T}}, \mathrm{and}\,\, \mathbf{g}=[g_{1},\cdots, g_{m},\cdots, g_{M}],
\end{equation}
where $h_{m} =\delta_m e^{-j \upsilon_m}$ and $g_{m} =\zeta_m e^{-j \varrho_m}$ denote the fading coefficients of the incident and reflection channels of the $m$-th element, respectively. Further, $\delta_m$ and  $\zeta_m$ are the channel amplitudes, while $\upsilon_m$ and $\varrho_m$ are the phase shifts.  All channels are assumed to be independent and identically distributed, and the Rician factor of the channels is denoted by $\mathcal{K}$. Therefore, the received SNR per user is given as follows:  
\begin{align}\label{before_coh}
\gamma^{\RIS}_k = \frac{P_k|\sum_{i=1}^{M} \delta_m \zeta_me^{-j \qty(\theta_m+\upsilon_m+\varrho_m)}|^2}{\sigma^2\mathscr{L}_1(|\boldsymbol{\psi}_0 - \boldsymbol{\psi}_1|)\mathscr{L}_2(|\boldsymbol{\psi}_0 - \boldsymbol{\psi}_2|)},
\end{align}
where $\theta_m$ denotes the phase shift introduced by the $m$-th RIS element. In \eqref{before_coh}, $\mathscr{L}_k(|\boldsymbol{\psi}_0 - \boldsymbol{\psi}_k|)$ is the path-loss between $U_k$ and the RIS that  is function of the distance in-between. We adopt the widely used Saleh-Valenzuela (SV) channel model \cite{sv1,
sv2} for modeling both links as follows:
\begin{equation}
\mathscr{L}_k^{[\mathrm{dB}]}(|\boldsymbol{\psi}_0 - \boldsymbol{\psi}_k|) = a + 10b \log_{10}(|\boldsymbol{\psi}_0 - \boldsymbol{\psi}_k|) + \xi,
\label{eq:pathloss}
\end{equation}
where $a$ is the path loss constant offset value, $b$ is the path loss attenuation constant, and $\xi$ is the shadow fading component with shadowing standard deviation set to $\sigma_{SF} = 0$ \cite{clustering_TDMA}.
After applying the coherent phase-shift design \cite{Basar2019}, i.e., setting $\theta_m = -(\upsilon_m + \varrho_m)$ in each user's time slot to maximize the channel gain, the SNR per user can be expressed as:
\begin{align}
\gamma^{\RIS}_k = \frac{P_k|\sum_{i=1}^{M} \delta_m \zeta_m|^2 }{\sigma^2\mathscr{L}_1(\sqrt{x^2_1+y^2_1+h^2})\mathscr{L}_2(\sqrt{x^2_2+y^2_2+h^2})}.
\end{align}
Therefore, the average spectral efficiency can be obtained as
\begin{align}\label{ach1}
\mathcal{R}^{\RIS}_k= \mathbb E\qty[\frac{1}{2}\log_2 \qty(1 +\gamma^{\RIS}_k )].
\end{align}
By using Jensen’s inequality, a lower bound of the spectral efficiency  in \eqref{ach1} is given by
\begin{align}
\mathcal{R}^{\RIS}_k \geq \mathcal{\Tilde{R}} = \frac{1}{2}\log_2 \qty(1 + \e\qty[\gamma^{\RIS}_k]).
\end{align}
The randomness in the SNR arises from the fading channels $\delta_m$ and $\zeta_m$ as well as and the random user coordinates $x_k$ and $y_k$. To facilitate a tractable analysis and derive a closed-form expression, we assume the users are located at the center of the coverage area $A_k$. Under this assumption, $\mathscr{L}_1(|\boldsymbol{\psi}_0 - \boldsymbol{\psi}_1|)$ and $\mathscr{L}_2(|\boldsymbol{\psi}_0 - \boldsymbol{\psi}_2|)$ become $\mathscr{L}\qty(\sqrt{\qty(d+\frac{L}{2})^2+h^2})$. Hence, a closed-form expression for the SNR can be obtained as follows:
\begin{align}\label{ana_ris}
\scalemath{0.9}{\mathcal{R}^{\RIS}_k = \log_2
\qty(1 + \frac{P\qty[M+\qty(\frac{\pi^2 L_{\frac{1}{2}}(-\BK)^4}{16(\BK+1)^2})\qty(M^2-M)]}{\sigma^2\mathscr{L}\qty(\sqrt{\qty(d+\frac{L}{2})^2+h^2})^2}).}
\end{align}

\section{Energy Efficiency}
This section analyzes the energy efficiency of pinching antenna and RIS systems by developing a power consumption model for the studied communication systems. The energy efficiency is defined as the ratio of the achievable data rate to the total power consumption of the system, measured in bits per joule (b/J), as follows:
\begin{equation}
EE = \frac{B\cdot\mathcal{R}^{\GC}}{\mathcal{P}_{\tot}},
\end{equation}
where $B$ denotes the transmission bandwidth in Hz, $\mathcal{R}^{\GC}$ denotes the sum spectral efficiency achieved by the two users.
$\mathcal{P}_{\tot}$ is the total power consumed by the whole communication system. In the following, we provide the energy efficiency analysis for the two communication scenarios.
\subsection{Pinching Antenna System}
Recall that, in the considered pinching antenna system, each user transmits with power $\frac{P_k}{2}$, while the relay transmits with a total power of $P_k$ to serve both users. Define $\nu \in (0, 1]$ as the power amplifier efficiency of the relay \cite{Bjornson2020}. Accordingly, the overall energy efficiency for the pinching antenna system can be given as follows:
\begin{align}
EE^{\PIN}=\frac{ B\cdot\sum^{\smallK}_{k}R_k^{\PIN}}
{
P_k/\nu + P_{\mathrm{RE}} +\sum^{\smallK}_{k}\qty(P_k/2 + P_{\mathrm{UE},k}) },
\end{align}
where $P_{\mathrm{RE}}$ and $P_{\mathrm{UE},k}$ denote the static power consumption of the relay equipment and the user equipment, respectively. 
\subsection{RIS-Assisted System}
The overall energy efficiency of RIS-assisted communication system can be expressed as follows:
\begin{align}
EE^{\RIS} & =\frac{ B\cdot\sum^{\smallK}_{k}R_k^{\RIS}}
{ P_{\RIS} + \sum^{\smallK}_{k}\qty(P_k + P_{\mathrm{UE},k})},
\end{align}
where $P_{\mathrm{RIS}}$ is the power dissipated by the RIS, which depends on the phase-shifter
of the individual elements. Hence, $P_{\RIS}$ can be given as follows:
\begin{align}
P_{\RIS}=MP_{\text{ph-sh}},
\end{align}
where $P_{\text{ph-sh}}$ is the power consumption of each phase shifter.

\begin{table}[t]
\caption{\vspace{-1.0mm}Simulation Parameters}
\label{table1}
\centering
\renewcommand{\arraystretch}{1.5} 
\scalebox{0.9}{\begin{tabular}{|c|c|}
\hline
\bf Parameter & \bf Value  \\[-0.6mm]
\bottomrule[1pt]
Carrier frequency  & $f_c=28$ GHz \cite{pin_per} \\ \hline
Bandwidth  & $B=1$ GHz \\\hline 
Waveguide/RIS height  & $h=3$ m \cite{pin_per}\\\hline
Distance from origin to user's region   & $d=25$ m \\\hline 
Side length of user's region    & $L=10$ m \\\hline
Equipment power consumption   & $P_{\mathrm{RE}}= P_{\mathrm{UE},k} =10$ dBm \cite{samy_OJ}  \\ \hline
Amplifier efficiency  & $\nu=0.5$ \cite{Bjornson2020}\\ \hline
path-loss parameter $a$ & $a=61.4$ \cite{sv1}\\ \hline
path-loss parameter $b$ & $b=2$ \cite{sv1}\\ \hline
RIS impairment severity  & $\epsilon=0.5$ \cite{phase_nosie1}\\\hline
Rician factor  & $\mathcal{K}=10$\\ \hline
Noise power  & $\sigma=-80$ dBm \\\hline
Phase-shifter consumption at mmWave    & $P_{\text{ph-sh}}=17.5$ mW \cite{mmwave_phase} \\[-0.6mm]
\bottomrule[1pt]
\end{tabular}}
\vspace{-3mm}
\end{table}
\section{Numerical Results}\label{sec:numerical}
In this section, we conduct simulations to assess the performance of pinching antenna and RIS-assisted systems. To ensure accuracy, we employ Monte Carlo simulations with $10^4$ realizations. We also verify the derived expression for the pinching antenna and RIS communication scenarios. Unless stated otherwise, the simulation parameters used are provided in Table \ref{table1}. 
\subsection{Hardware Impairments for Pinching Antennas and RIS}

In our comparison, we further investigate the impact of hardware impairments on the performance of pinching antenna and RIS-assisted systems. To enrich the analysis, the following hardware impairments are considered 
\begin{enumerate}
    \item \textbf{Lossy pinching antenna:} In practice, the available transmit power $P_k$ may not be fully emitted by the pinching antenna, leading to transmit power loss \cite{pin_per}. This effect is modeled by $\beta P_k$, where $\beta \in [0,1]$ represents portion of the transmit power emitted by the pinching antenna.

    \item \textbf{Lossy waveguide with end-feeder:} The dielectric waveguide exhibits a propagation loss of approximately $0.08$ dB/m \cite{pozar2000microwave}, which can significantly affect signal strength over long distances. Moreover, a single feed point is located at one end of the waveguide is considered, where we evaluate the performance of the user located at the far end of the waveguide feeder.

    \item \textbf{Lossy waveguide with mid-feeder:} This scenario is similar to Case 2 in terms of propagation loss, however instead employs two co-located feed points at the center of the waveguide \cite{pin_per}, which can send signals bidirectionally along the waveguide depending on the served user’s location.

   \item \textbf{RIS with phase noise per element:} Recall that $\theta_m$ denotes the optimal phase shift required for coherent beamforming. However, due to hardware imperfections, the actual phase-shift applied by each element is $\tilde{\theta}_m = \theta_m + \hat{\theta}_m$, where $\hat{\theta}_m \sim \mathcal{U} \qty[ -\epsilon\pi, \epsilon\pi ]$ represents the phase noise \cite{phase_nosie1,phase_nosie2}, $\mathcal{U}$ denotes the uniform distribution and $\epsilon$ measures
the severity of the impairments at the RIS. Note that, $\epsilon=0$ corresponds to perfect phase adjustment (i.e., without phase noise), while $\epsilon=1$ represents the worst-case scenario.
\end{enumerate}
\begin{figure}[t]
    \centering
    \includegraphics[width=2.9in]{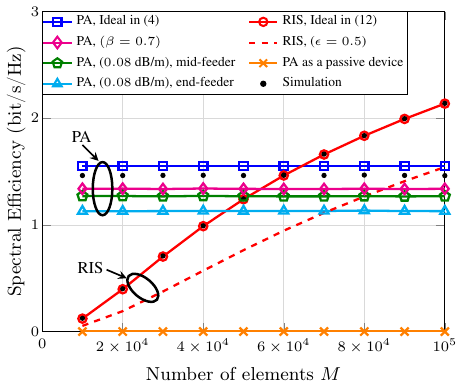}
\vspace{1.0mm}
\renewcommand{\figurename}{\footnotesize Fig.}
\captionwidth
\caption{\footnotesize  Spectral efficiency of pinching antennas (PA) and RIS-assisted systems with transmit power $p=15$ dBm.}
\label{spec_comp}
\vspace{-5mm}
\end{figure}
\subsection{Spectral Efficiency}
In Fig. \ref{spec_comp}, the spectral efficiency is plotted versus the number of reflecting elements of the RIS and compared with the performance of pinching antenna systems, considering the hardware impairments for both technologies as discussed earlier. It can be observed that as the number of RIS elements increases, the impact of hardware impairments becomes more prominent, indicating that large-scale RIS deployments may suffer from reduced efficiency due to these imperfections. For pinching antenna systems, the performance degradation caused by mid-feeder configuration and antenna loss case remains relatively close to the simulation curve of the ideal case. As expected, the worst performance is observed when the waveguide uses a single feeder located at one end, due to higher propagation losses. Interestingly, under this worst-case pinching antenna scenario, the RIS with phase-noise impairments requires approximately $7 \times 10^4$ reflecting elements to outperform pinching antenna setup. This number increases to around $8 \times 10^4$ elements to outperform the mid-feeder configuration and antenna loss case. Additionally, as discussed earlier, using the pinching antenna in a similar manner to RIS to perform communication over two time slots yields no performance gains. Finally, a good match between the closed-form expressions and the simulation results is achieved.

\begin{figure}[t]
    \centering
    \includegraphics[width=2.9in]{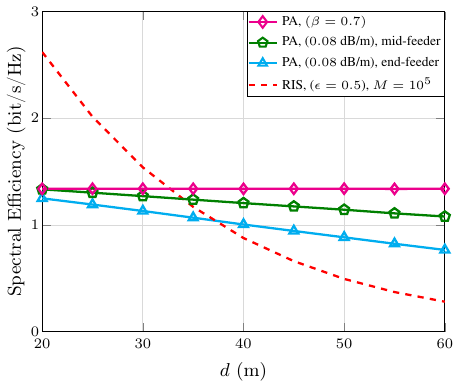}
\vspace{-2.0mm}
\renewcommand{\figurename}{\footnotesize Fig.}
\captionwidth
\caption{\footnotesize  Spectral efficiency versus the distance from origin to user's region with transmit power $p=15$  dBm.}
\label{distances}
\vspace{-1mm}
\end{figure}

In Fig. \ref{distances}, we investigate the spectral efficiency performance of pinching antennas and RIS when the horizontal distance between user’s service regions increase. As expected, the single-feeder waveguide exhibits a sharper decline in performance compared to double-feeder waveguide as $d$ increase.
One can also observe that RIS is more susceptible to severe path loss, where the performance gap between the pinching-antenna and RIS increases significantly with $d$. This result shows the great flexibility of pinching-antenna systems to combat path loss, and demonstrates greater robustness to distance variation compared to RIS.

\subsection{Energy Efficiency}

\begin{figure}[t]
    \centering
    \includegraphics[width=2.9in]{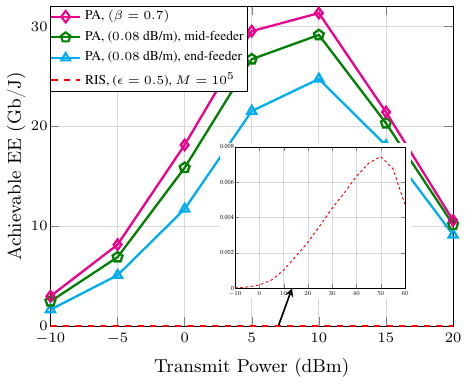}
\vspace{-2.0mm}
\renewcommand{\figurename}{\footnotesize Fig.}
\captionwidth
\caption{\footnotesize  Energy efficiency versus transmit power in  dBm.}
\label{EE_HW}
\vspace{-1mm}
\end{figure}
Fig. \ref{EE_HW} illustrates the system energy efficiency as a function of transmit power (in dBm) for both the pinching antenna and RIS-assisted systems, considering hardware impairments. The results clearly demonstrate that the RIS-assisted system becomes highly inefficient in the mmWave band when compared to the significantly superior performance of the pinching antenna system, as illustrated in the figure. For instance, the pinching antenna system achieves peak energy efficiency at $10$ dBm, reaching more than $32$ Gb/J. In contrast, RIS-assisted system peaks at $0.007$ Gb/J, and at a much higher transmit power of $50$ dBm, which is not practical in many wireless communication scenarios. 
This performance deterioration of RIS is due to the need for a large number of elements (in the order of $10^4$) to overcome the severe double path loss in mmWave frequencies.
%
%

%

%
\section{Conclusions}\label{sec:conc}
We have compared the emerging pinching antenna systems with RIS-assisted systems in mmWave bands. The key observation is that RIS requires massive number of elements to be competitive with pinching antenna systems in terms of spectral efficiency, which severely impact the energy efficiency performance of RIS. Additionally, pinching antenna systems demonstrated greater robustness against hardware impairments and severe path loss, especially when the waveguide feeder point is placed at the center of the waveguide.
%
An important direction for future research is the study of the coexistence between pinching antennas and RIS, which may offer complementary capabilities in next-generation wireless networks.

\vspace{0mm}

\linespread{1.05}
\vspace{0mm}
\sloppy
\bibliographystyle{IEEEtran}
\bibliography{IEEEabrv,References}
\pagestyle{empty}

\end{document}